\documentstyle[aps,epsfig,amssymb]{revtex}
\newcommand{\mel}[3]{\langle #1 | #2 | #3 \rangle }
\newcommand{\ket}[1]{| #1 \rangle }

\newcommand{\iu }{{\rm i}}
\newcommand{\tr}{\rm tr}
\begin{document}
\title{Controlling  decoherence of a two-level-atom in a lossy cavity}
\author{M.\ Thorwart, L.\ Hartmann, I.\ Goychuk\footnote{on leave of absence 
from Bogolyubov Institute for Theoretical Physics, Kiev, Ukraine}, 
and P.\ H\"anggi}
\address{
              Institut f\"ur Physik -
              Universit\"at  Augsburg, Universit\"atsstr.\ 1, D - 86135
Augsburg,
              Germany\\
}
\date{Date: \today}
\maketitle
\begin{abstract}
By use of external periodic driving sources, we demonstrate the possibility of
controlling the  coherent as
well as the decoherent dynamics of
a two-level atom placed in a lossy cavity.
 The control
of the coherent dynamics is elucidated for the phenomenon of {\em coherent
destruction of tunneling} (CDT),
 i.e., the coherent dynamics of a driven two-level atom
in a quantum superposition state can be brought practically to a complete
standstill. We study this phenomenon for different initial preparations of the
 two-level atom. We then proceed to investigate  the  decoherence
originating from
the interaction of the two-level atom with a lossy cavity mode. 
The loss mechanism is described in terms of a microscopic model 
that couples the cavity mode to a bath of harmonic field modes. 
A suitably tuned external cw-laser field applied to the
two-level atom
slows down considerably the  decoherence of the atom. We
demonstrate the suppression of decoherence for
two opposite initial preparations of the atomic state:
a quantum superposition state as well as the ground state.
These findings can be used to the effect of a proficient battling of
decoherence
in qubit manipulation
processes.
\\
\end{abstract}
\vspace{-2mm}
%
%
\section{Introduction}
The idea of controlling the coherent dynamics of a quantum system
by an external time-dependent force has found
 wide spread experimental and theoretical interest in many areas of physics
(for  reviews see
\cite{Grifoni98,Shapiro,Tannor}).
It is, e.g., a commonly used tool
to manipulate trapped atoms in quantum optics \cite{WallsBook,Nobelprize97}
as well as to control chemical reactions by a strong laser field
\cite{Grifoni98,Shapiro,Tannor}.
In the context of quantum optics, it has been demonstrated experimentally
\cite{Noel98} that a frequency-modulated excitation of a two-level
atom by use of a microwave field driving transitions between two Rydberg Stark
states of potassium significantly modifies the time evolution of the system.
 In the context of tunneling systems it has also  been demonstrated
that it is in principle  possible to completely
suppress the coherent tunneling of an initially localized wave packet
in a double-well potential by an external, suitably designed time-periodic
cw-perturbation
({\em coherent destruction of tunneling}) \cite{Grossmann92a}.

However, real quantum systems are always in contact with
their environment. The coherent dynamics is then  usually destroyed due
to the influence of the large number of environmental degrees of freedom.
Not only the phase of the quantum system is dis\-turbed (decoherence)
\cite{JoosBook} but also
energy exchange (dissipation) \cite{Weiss93,Knight98}
takes place between the system under consideration and
the environment. An example of such a system-bath interaction is the
ensemble of electromagnetic field modes in a cavity,
each of which is described as a
quantum mechanical harmonic oscillator \cite{WallsBook}.
Each mode interacts with
 an atom trapped in the cavity. On the other side, the cavity modes themselves
 are also not isolated from the macroscopic environment; as such they are more
 realistically described as damped quantum harmonic oscillators.
 A topic of fundamental interest is the decay of quantum superpositions of
states.
 In \cite{Garraway98} it is shown how quantum optical nonclassical states
 are highly sensitive to dissipation stemming from a zero-temperature
 heat bath. Experimental works studying decoherence systematically are rare.
 In \cite{Haroche98} the decoherence of mesoscopic superpositions
 of field states in the cavity has been investigated.
 In a recent work, Wineland and collaborators \cite{Wineland00}
 demonstrate that the decoherence
 rate scales with the square of a quantity that describes the separation
between
 two initial  states. Moreover, Knight and co-workers \cite{Knight99}
 proposed an experimental
 scheme to probe the decoherence of a macroscopic object.

In this spirit, the question arises to which extent it is possible to
control the dynamics of a quantum system in presence of decoherence and,
moreover,
whether the effect of decoherence can be minimized
by an external time-dependent force, e.g., by a laser field
\cite{Grifoni98,Grossmann92a,Grossmann92,Oelschlaegel,Lloyd99,Vitali99,Agarwal00}. 
 To achieve this goal, various approaches have been undertaken in recent
years. (i) It
has been shown that  the effect of coherent destruction of tunneling (see
above) can be
used  to {\it slow down} the relaxation of a quantum system to its asymptotic
equilibrium
\cite{Oelschlaegel}. (ii) Moreover, a suitable tailored sequence of
radio-frequency
pulses (``quantum bang-bang'' \cite{Lloyd99} or ``parity kicks''
\cite{Vitali99}) that repeatedly flip the state of a two-level atom
 may be used to suppress decoherence. 
 (iii) The cavity-induced spontaneous emission of a two-level atom can 
 be manipulated by a strong rf field which couples to the cavity mode 
 \cite{Agarwal93}. 
 (iv) The manipulation of the system-bath
interaction
 by a fast frequency modulation also results in slowing down decoherence
 and relaxation \cite{Agarwal00}.

%
The objective of this work is to study the influence of a time-periodic 
driving field on the dynamics of a two-level atom. In the first 
part of this work (section \ref{nodissipation}), 
we deal with the objective to ``freeze'' a coherent dynamics, i.e., we 
shall employ an effect known as coherent destruction of 
tunneling. Most importantly, we investigate this freezing phenomenon 
from the viewpoint of its dependence on different initial preparations. 

In the second part of this work (section \ref{dissipation}), 
we do not elaborate further on  
 the effect of coherent destruction tunneling, but instead investigate the 
 control of decoherence of a two-level atom placed in a lossy cavity. 
 Our model consists of  a two-state system 
which is coupled to a time-dependent periodic field. 
 The driven two-state system interacts furthermore 
 with one mode of the cavity having the frequency $\Omega$. This 
 mode is itself 
 damped by the coupling to a bath of harmonic oscillators
({\em lossy cavity}).
It is known \cite{Garg85} that a Hamiltonian consisting
of (1) a system part, (2) a harmonic oscillator with frequency
$\Omega $ that is
being  coupled to the system, and (3) a bath of
harmonic oscillators which are coupled to this very single harmonic oscillator
can be mapped onto a
Hamiltonian composed of the system part coupled to a harmonic bath
with an effective
spectral density. This effective spectral density possesses a Lorentzian-shaped
 peak at $\Omega$.
The completely isolated atom (no driving, no cavity) evolves in time 
in a coherent way according to the Schr\"odinger equation. 
It is this dynamics which we want to preserve and protect as far as possible 
from  the decoherent influence of the environment. 
Our major finding is that a cw-control field can indeed be used to (i) 
reduce decoherence and (ii) to restore to some extent the unperturbed,  
non-dissipative time-evolution.
%

%
%
\section{The driven two-level atom}
\label{nodissipation}
\subsection{Floquet Formalism}
To start we consider a Hamiltonian describing a
two-level atom with the ground state $\ket{1}$ and
an excited state $\ket{2}$. The energy levels are separated by the
energy $\hbar \Delta_0$. The atom with the transition dipole
moment $\mu$ is driven within the long wavelength approximation by an
external,
time-dependent laser field of the form  ${\cal E}(t)={\cal E}_0 \,
\cos(\omega_{\rm L}
t)$
 with frequency $\omega_{\rm L}$ and amplitude
${\cal E}_0$, yielding the driven quantum system
\begin{equation}
H(t) = -\frac{\hbar}{2} [\Delta_0 \hat{\sigma}_z + s(t)
\hat{\sigma}_x]  \, . \label{drivtls}
\end{equation}
Here, the matrices   $\hat{\sigma}_i, \, i=x,y,z$ denote the Pauli spin
matrices.
The part involving $s(t) = s \, \cos(\omega_{\rm L} t)$ with $s=2\mu{\cal
E}_0 / \hbar$
presents  the time-dependent driving which couples
to the transition dipole moment $\mu$ of the atom.
 Note that within this scaling the amplitude $s$
possesses the dimension of a frequency.
The driven time evolution of the populations of the
energy levels exhibits an oscillatory behaviour. For an
initial preparation of the atom in the ground state and for a resonant
driving,
i.\ e.\  $\omega_{\rm L}= \Delta_0$, and with $s$ not large we can invoke the
rotating wave
approximation. The population of  each state then oscillates  between 0
and 1 with the Rabi frequency $\Omega_{\rm R} = s /2$.
Because the Hamiltonian (\ref{drivtls}) is periodic in time
with the period ${\cal T} = 2 \pi / \omega_{\rm L}$, i.e.,
$H(t+{\cal T}) = H(t)$, we next apply for the general case away from
resonance the Floquet formalism
\cite{Grifoni98}.  The time-dependent Schr\"odinger equation may be written as
\begin{equation}
\{ H(t) - \iu  \hbar \partial/\partial t \} \ket{\psi(t)} = 0.
\label{schr}
\end{equation}
According to the Floquet theorem, there exist solutions to eq.\  (\ref{schr})
of the form
\begin{equation}
\ket{\Psi_{\alpha}(t)} = \exp (-\iu \varepsilon_{\alpha}t/\hbar)
\ket{\Phi_{\alpha}
(t)} \, , \label{floquet1}
\end{equation}
with $\alpha = 1, 2$.
The periodic function $\ket{\Phi_{\alpha}(t)}$ are termed the Floquet
modes and these obey
\begin{equation}
\ket{\Phi_{\alpha}(t+{\cal T})} = \ket{\Phi_{\alpha}(t)} \, .
\end{equation}
Here, $\varepsilon_{\alpha}$ is the so called Floquet characteristic exponent
or the {\em quasienergy}, which is {real-valued} and unique up to multiples
of $\hbar \omega_{\rm L}$. Upon substituting eq.\ (\ref{floquet1}) into
the Schr\"odinger equation (\ref{schr}) one obtains the eigenvalue equation
for the quasienergy $\varepsilon_{\alpha}$
\begin{equation}
{\cal H}(t)\ket{\Phi_{\alpha}(t)} =  \varepsilon_{\alpha}
\ket{\Phi_{\alpha}(t)}\,
\end{equation}
with the Hermitian operator
\begin{equation}
{\cal H}(t) \equiv H(t)- \iu  \hbar \partial/\partial t \, .
\label{hamcal}
\end{equation}
We stress that the Floquet modes
\begin{equation}
\ket{\Phi_{\alpha'}(t)} = \ket{\Phi_{\alpha}(t)} \exp (\iu n \omega_{\rm L} t)
\equiv \ket{\Phi_{\alpha n}(t)}
\end{equation}
with $n$ being an integer number $n=0, \pm 1, \pm 2, \dots$ yield equivalent
solutions to  eq.\ (\ref{floquet1}) but with the shifted quasienergy
\begin{equation}
\varepsilon_{\alpha} \rightarrow \varepsilon_{\alpha'} = \varepsilon_{\alpha}
+ n \hbar \omega_{\rm L} \equiv \varepsilon_{\alpha n} \, .
\end{equation}
Therefore, the index $\alpha$ corresponds to a whole class of solutions
indexed
by $\alpha' = (\alpha , n)$. The eigenvalues $\{ \varepsilon_{\alpha} \}$
 can thus be mapped into a first Brillouin zone obeying
$-\hbar \omega_{\rm L} / 2 \le \varepsilon <  \hbar \omega_{\rm L} / 2$.
It is clear that for our choice of the external driving force, i.e.\
$s(t) = s \, \cos \omega_{\rm L} t$ the quasienergies are functions of
the driving amplitude $s$ and the driving frequency $\omega_{\rm L}$. For
adiabatically vanishing external driving they merge into the
eigenvalues of the time-independent part of the Hamiltonian (\ref{drivtls}),
i.e.,
\begin{equation}
\varepsilon_{\alpha n} (s,\omega_{\rm L})
\stackrel{s\rightarrow 0}{\longrightarrow} \mp \hbar \Delta_0 / 2
+ n \hbar \omega_{\rm L}\, ,
\end{equation}
where the negative (positive) sign corresponds to $\alpha=1 \, (\alpha=2)$.
The Floquet modes, correspondingly,
 turn into the eigenfunctions $\ket{\alpha}$ multiplied
 by an additional phase factor, i.e.,
\begin{equation}
\ket{\Phi_{\alpha n}(t)} \stackrel{s\rightarrow 0}{\longrightarrow}
\ket{\alpha} \exp(\iu \omega_{\rm L} n t)\, .
\end{equation}
For a finite driving strength $s\ne 0$, the determination of the quasienergies
$\varepsilon_{\alpha}$ requires the use of numerical methods. The interested
reader
is referred in this context to the literature \cite{Dittrich97,Shirley65}.
However,
we here state without proof  that in the
high-frequency regime $\Delta_0 \ll {\mbox{max}}[\omega_{\rm L},
(s \, \omega_{\rm L})^{1/2}]$ the difference between the two quasienergies
is given by
\cite{Grossmann92}
\begin{equation}
\varepsilon_{2,-1}-\varepsilon_{1,1} = \hbar \Delta_0 J_0(s/\omega_{\rm L})\,
\label{cdtbessel},
\end{equation}
where $J_0$ denotes the zeroth-order Bessel function of the first kind.

\subsection{Freezing the coherent dynamics of a driven two-level system}
Eq.\  (\ref{cdtbessel}) implies a most interesting consequence 
for a driven two-level system
\cite{Grossmann92}:  If one chooses the driving parameter $s$ and
$\omega_{\rm L}$ in
such a way  that the argument of the Bessel function is at a zero of
the Bessel function, the
splitting between the quasienergy vanishes. Possible
transitions between the
Floquet states are then at most induced by the remaining periodic
time-dependent parts
of  the corresponding Floquet modes $\ket{\Phi_{\alpha}(t)}$.
This effect has been discovered in the context of
tunneling systems.  There, a wave packet being an equally weighted
superposition of the symmetric and antisymmetric
ground state is initially
localized at one side of a double-well potential. By applying an external
suitably
tailored  periodic field, the wave packet can be stabilized and
can be prevented from coherently tunneling back and forth
 between the two wells, i.e.,   one finds
 {\em coherent destruction of tunneling} (CDT) \cite{Grifoni98,Grossmann92a}.
We emphasize here that the crossing of two tunneling related quasienergy
levels yields a
{\em necessary}
 (but not sufficient) criterion for the suppression of coherent
 tunneling \cite{Grossmann92}.

The challenge
we want to address next is as follows: How does the driven dynamics of
a two level  atom  that is being prepared in some arbitrary initial
state evolve when the corresponding condition for the parameters
obey the CDT-condition  (\ref{cdtbessel})? The system dynamics can be described
by its density operator $\hat{\rho}(t)$, which is a $2 \times 2$-matrix, 
i.e.,  

\begin{equation}
\hat{\rho}(t) =  \hat{I}/2 + \sum_{i=x,y,z}\sigma_i(t)\hat \sigma_i/2 \, ,
\label{twobytwo}
\end{equation}
where the expectation values
$\sigma_i(t) := {\rm Tr}\{\hat \rho(t)\hat \sigma_i\}, i=x,y,z$
are the dynamical quantities of interest.  $\hat I$
denotes the unit matrix and $\sigma_x(t)$ and $\sigma_y(t)$ are related to the
coherences (the off-diagonal elements) of $\hat{\rho}(t)$ while
$\sigma_z(t)$ is the population difference between the two energy eigenstates
$\ket{\alpha}$.
This implies that the state of the quantum system at time $t$
is fully determined
by the knowledge of the three expectation values $\sigma_i(t)$.
%

To determine the state of the driven two-level system at time $t$, we
consider the Heisenberg equation of motion for the density matrix.
Using the commutation relations for the $\hat\sigma_i$, we  arrive
at the equation of motion for the expectation values $\sigma_i(t)$ in
(\ref{twobytwo}), i.e.,
\begin{eqnarray}
\dot \sigma_x(t) &=& -\Delta_0 \sigma_y(t)\, , \nonumber \\
\dot \sigma_y(t) &=& \Delta_0 \sigma_x(t) - s(t) \sigma_z(t)\, , \nonumber \\
\dot \sigma_z(t) &=&  s(t) \sigma_y(t)\, . \label{heisi}
\end{eqnarray}
To study the dependence of the effect of coherent
destruction of tunneling on the initial
preparation we first choose as initial state an equally weighted
coherent superposition of the two unperturbed energy eigenstates, i.e.,
\begin{equation}
\ket{\Psi (t=0)} = \frac{1}{\sqrt{2}} (\ket{1} + \ket{2}) \, ,
\label{inistate}
\end{equation}
corresponding to $\sigma_x(t=0)=1, \sigma_y(t=0)=\sigma_z(t=0)=0$.
We solve the set of coupled differential equations (\ref{heisi}) numerically
by a standard fourth order Runge-Kutta integration algorithm with
adaptive step-size control.
In Fig.\ \ref{fig1}a, the time-dependence of the three expectation values
is depicted. The driving parameters are chosen such that the condition
(\ref{cdtbessel}) is fulfilled: in doing so we use \  $\omega_{\rm
L}=50 \Delta_0$
and
$s=120.241... \Delta_0$. Surprisingly  {\it all three}
expectation values  $\sigma_i(t)$ can be
brought simultaneously to an almost perfect standstill !

Next, we choose  the {\em ground state} as initial state, i.e., we use
$\ket{\Psi (t=0)} =
 \ket{1}$. This corresponds to $\sigma_x(t=0)=\sigma_y(t=0)=0,
\sigma_z(t=0)=1$.
 The result is depicted in Fig.\ \ref{fig1}b.
  Applying to the so prepared two-level system a
 laser field obeying the CDT-condition (\ref{cdtbessel}), 
 we find that the y-component $\sigma_y(t)$
 and the z-component $\sigma_z(t)$ exhibit
 strong  oscillations.  This oscillations
follow from the numerically evaluated
 Floquet theory for the driven two-level system, and are {\em not}
described by the Rabi-oscillations as predicted from a rotating wave
approximation; this latter approximation is strongly violated
for our chosen set of driving parameters. In contrast, $\sigma_x(t)$ can be
 stabilized around the initial value of zero. This finding is in accordance
 with the CDT-phenomenon: it reflects the fact that the corresponding two
equally weighted ("left" and "right")
localized parts of the ground state wave function of a double-well potential, 
as represented within a
localized representation, can each be stabilized too.

This CDT-effect opens the doorway to manipulate the influence of an environment
on a quantum system. It is  known \cite{Oelschlaegel}
that the coherent destruction of tunneling  survives to some extent
 in presence of a coupling to the environment. Certainly, the system will
 relax in presence of an environment; however, as it is
 demonstrated in \cite{Oelschlaegel}, the relaxation process
 can  considerably be slowed down in the presence of a CDT-field.

In view of using differing initial preparations, the following remark
should be made. From the  viewpoint of stabilizing
the state of a qubit (characterized by a
quantum
mechanical two-level system) in a quantum
information processor \cite{quantComp}, it is of foremost interest to
stabilize
the coherent superposition of two states of the qubit. Thus, our
first choice (\ref{inistate}) is of relevance in the context
of the possibility for quantum computing.
Moreover, fundamental questions concerning the decoherence
of superposition states arise for the physics  that occurs
 when one crosses the interface between the  classical 
and  quantum world, and vice versa
\cite{JoosBook,Wineland00}.

\section{Control of decoherence for a two-level atom}
\label{dissipation}
In this section we shall study the influence of an applied cw-control field 
for reducing decoherence of a two-level atom placed in a lossy cavity. 
\subsection{Driven two-level atom in a lossy cavity}
%
To start we consider a two-level atom
in a dissipative environment, e.g., a lossy cavity wherein the
leakage of photons damps the radiation field. Additionally,
the atom may be manipulated by a time-dependent external field
like a laser beam. In our model, the driven two-level atom is represented
by the Hamiltonian  (\ref{drivtls}). It is coupled to one mode of the
cavity which is described by one harmonic oscillator with
frequency $\Omega$,  characterized by the annihilation and creation
operators $\hat{B}$ and $\hat{B}^{\dagger}$ which fulfill the usual
commutation
relations for bosonic field operators. The coupling constant is denoted by $g$
and has the dimension of a frequency. This cavity mode
is damped by a bilinear coupling to a bath of
harmonic oscillators of
frequencies $\omega_i$. They are similarly described by  bosonic
annihilation and creation operators $\hat b_i$ and $\hat{b}_i^{\dagger}$.
 The coupling constants of the
cavity mode to the harmonic bath are given by $\kappa_i$ and have the
dimension
of a frequency.
The total system-bath Hamiltonian is therefore written as
\begin{eqnarray}
H(t) & = & -\frac{\hbar}{2} [\Delta_0 \hat{\sigma}_z + s(t)
\hat{\sigma}_x]  \nonumber \\
& & \mbox{} + \hbar \Omega (\hat{B}^{\dagger}\hat{B}+ \frac{1}{2})
+\hbar g (\hat{B}^{\dagger}+\hat{B}) \hat{\sigma}_x  \nonumber \\
& & \mbox{} + \sum_{i=1}^N \hbar \omega_i (\hat{b}_i^{\dagger}\hat{b}_i+
\frac{1}{2})
+ \hbar (\hat{B}^{\dagger}+\hat{B}) \sum_{i=1}^N \kappa_i
(\hat{b}_i^{\dagger}+\hat{b}_i) \, .
\label{dissdrivtls}
\end{eqnarray}
The influence of the bath on the
two-level atom plus cavity mode is 
fully characterized by the spectral density
\begin{equation}
J(\omega)=2\pi \sum_{i=1}^N \kappa_i^2 \, \delta(\omega - \omega_i).
\end{equation}
We let the number of bath modes going to infinity ($N\rightarrow \infty$)
and choose
an Ohmic spectral density for the bath oscillators with an
exponential cut-off at some large
frequency $\omega_c\gg \Delta_0, \omega_{\rm L},\Omega$, i.e.,
\begin{equation}
J(\omega)=\frac{2\Gamma}{\Omega} \omega \exp(-\omega / \omega_c) \, ,
\label{ohmic}
\end{equation}
Here, we have introduced the damping constant $\Gamma$ which is related
to the quality factor of the cavity.  Since the cavity mode as well
as the bath oscillators are described by harmonic oscillators,
we follow the approach in \cite{Garg85} and map the Hamiltonian
(\ref{dissdrivtls}) onto a Hamiltonian where the
central system, i.e., the two-level atom, is now bilinearly
coupled to a bath of
{\em mutually non-interacting} harmonic oscillators with an effective
spectral density $J_{\rm eff}(\omega)$. Upon letting the cut-off frequency
going to infinity, i.e.,
$\omega_c \rightarrow \infty$, this effective spectral density emerges as
\begin{equation}
J_{\rm eff}(\omega) =
\frac{16 \Gamma}{\Omega}
\frac{
g^2 \omega \Omega^2}
{
(\Omega^2-\omega^2)^2 + 4 \omega^2 \Gamma^2
} \, . \label{jeff}
\end{equation}
 For small
frequencies $\omega$, it increases linearly like in the original Ohmic
spectral density $J(\omega)$. However, it has a Lorentzian shaped peak
at $\omega = \Omega$ with a line width $\Gamma<\Omega$.

In the following section, we make extensive use of
the bath autocorrelation function ${\cal M}(t) = {\cal M}'(t) + \iu
{\cal M}''(t)$,
which is obtained in terms
of the effective spectral density $J_{\rm eff}(\omega)$, i. e.
\begin{equation}
{\cal M}(t)={1 \over \pi}\int_0^\infty d\omega J_{\rm eff}(\omega ) \left [
\mbox{coth} \left ( {\hbar \omega \over 2 k_{\rm B} T} \right ) \cos (\omega
t) - i\sin( \omega t)\right ] \, .
\label{response}
\end{equation}
All our further considerations treat the case when the bath
is at zero temperature, i.e.\ $T=0$. In this limit, and for our
choice of the effective spectral density (\ref{jeff}), we
obtain for the real and imaginary part, respectively, the analytical results
\begin{eqnarray}
{\cal M}'(t) & = &\frac{16 \Gamma}{\pi \Omega} g^2 \Omega^2
\big[
\frac{\pi}{4} \frac{1}{\Gamma \sqrt{\Omega^2 - \Gamma^2}} e^{-\Gamma t} \,
\cos(\sqrt{\Omega^2 - \Gamma^2} t) -
\int_0^{\infty} dy \frac{y e^{-yt}}{(y^2+\Omega^2)^2- 4 y^2 \Gamma^2}
\big] \nonumber\, , \\
{\cal M}''(t) & = & - 4 g^2 \frac{\Omega}{
\sqrt{\Omega^2 - \Gamma^2}} e^{-\Gamma t} \sin(\sqrt{\Omega^2-\Gamma^2}t)\, .
\label{corrfunc}
\end{eqnarray}
The quantity of interest is the reduced density matrix for 
the two-level system which we denote --
just as
in the undamped case -- by $\hat\rho(t)$. It follows by
tracing over the bath degrees of freedom in the full
density operator $\hat{W}(t)$ which corresponds to the
system-plus-bath Hamiltonian
(\ref{dissdrivtls}), i.e., $\hat{\rho}(t) = \tr_{\rm B} \hat{W}(t)$.
Like in the
deterministic case in eq.\ (\ref{twobytwo}), $\hat{\rho}(t)$ is
fully characterized by the expectation values $\sigma_i(t), i=x,y,z$. We
shall determine
their corresponding equations of motions next.

\subsection{Bloch - Redfield Formalism}

To deal with quantum dissipative systems, several techniques have been
developed \cite{Grifoni98,Weiss93,Knight98,Dittrich97}.
A very efficient numerical algorithm for general quantum system
 with
a discrete eigenvalue spectrum has been developed by Makri and Makarov
within the real-time path-integral formalism \cite{QUAPI}.
It has also been
applied to spatially 
continuous tunneling systems in presence of driving \cite{Thorw}.
Moreover, the real-time path-integral formalism has extensively been used
to describe a moderate-to-strong (!) two-level system-bath interaction
\cite{Grifoni98,Weiss93,Garg85,Hartmann98b}. Recently, the former scheme has been 
 generalized to describe 
 multi-level, driven vibrational and tunneling dynamics in 
 \cite{Thorw00}. At weak system-bath coupling
 the Nakajima-Zwanzig projector operator theory \cite{Nakajima}
 provides a powerful tool
to describe the corresponding reduced density matrix dynamics.

For our quantum optical problem at hand, the suitable method of choice
in presence of a physically realistic  weak system-bath coupling is
  the projection
operator technique: it yields in Born approximation the
generalized master equation. It can be simplified
further
without loss of accuracy in leading order in the (weak) coupling strength
$g$ by invoking the Markovian approximation \cite{Hartmann00}.
For a strong harmonic driving this objective was formally (only) 
 developed a long time
ago by Argyres and Kelley \cite{Argyres}. Following the reasoning
in \cite{Argyres} (see in this context also \cite{Igor95}) we recently
 have derived
for this case of  a driven spin-boson problem with an {\it arbitrary}
control field
 the explicit set of  coupled, Bloch-Redfield type equations \cite{Hartmann00}
\begin{eqnarray}
\dot \sigma_x (t)&=& -\Delta_0\sigma_y (t)\, , \nonumber \\
\dot \sigma_y (t)&=& \Delta_0 \sigma_x(t) -
s(t)\sigma_z(t)-
\Gamma_{1}(t)\sigma_y(t)-\Gamma_{2}(t)\sigma_x(t)-A_y(t)
\,, \nonumber \\
\dot\sigma_z(t)&=&s(t)\sigma_y(t)
- \Gamma_{1}(t)\sigma_z(t)- \Gamma_{3}(t)\sigma_x(t)-A_z(t)\, .
\label{markov}
\end{eqnarray}
The time-dependent rates $\Gamma_{i}(t)=\int_{0}^{t}
dt'{\cal M}'(t-t')b_{i}(t,t')$ ,
together with the inhomogeneities
$A_y(t)={\rm Re}F(t)$, $A_z(t)={\rm Im}F(t)$,
with $F(t)=(1/2)\int_{0}^{t}dt' {\cal M}''(t-t')
[u^2(t,t')-v^2(t,t')]$
determine the dissipative action of  the thermal bath on the  two-level atom.
The functions  ${\cal M}'$ and ${\cal M}''$
denote the real part and imaginary part, respectively,
of the  correlation function ${\cal M}$ given in eqns.\ (\ref{corrfunc}).
The quantities
$ u(t,t')=\mel{1}{\hat{U}(t,t')}{1} + \mel{2}{\hat{U}(t,t')}{1}$ and
$v(t,t')=\mel{1}{\hat{U}(t,t')}{2} + \mel{2}{\hat{U}(t,t')}{2} $ are
sums of matrix elements of the time evolution
operator $\hat{U}(t,t')$ of the {\em isolated} (i.e., $g=0$)
driven two-level
system. The functions $b_{i}$ read
$b_{1}= {\rm Re} u v^*$,
$b_{2}=-(1/2)\;{\rm Im}(u^2-v^2)$, and $b_{3}=(1/2)\;{\rm Re}(u^2-v^2)$.
 Note that this set of equations is valid in the
parameter
regime  $g \ll \Delta_0/2$. One can demonstrate that
for the undriven case, i.e.,  $s=0$,  the
analytic solution of eq.\  (\ref{markov}) in first order in $g$
reproduces the
analytical path integral weak-damping results in Refs.\  \cite{Weiss93,Sigmax}.
%
\subsection{Controlling the decoherence of a quantum superposition of
states}  \label{results}
The idea of controlling the decohering influence of the environment on
a quantum system by an external time-dependent field is demonstrated for the
case of the
two-level atom which is initially prepared in an equally weighted superposition
of the two energy eigenstates given by 
$\sigma_x(t=0)=1, \sigma_y(t=0)=\sigma_z(t=0)=0$. In doing so, we consider four
different situations: (1) first, we look at the  isolated two-level atom
dynamics 
without driving  and without coupling the atom to the lossy cavity mode.
This case corresponds to setting  $s=0$ and
$g=0$. Case (2) is devoted to the driven two-level dynamics. 
  We switch on a coherent driving cw-field but keep the system
isolated from the bath, i.e. $s\ne0$ and
$g=0$. In case (3) we investigate how the undriven system dynamics
relaxes in presence of a dissipative coupling to the  bath. We therefore 
set $s=0$ and $g\ne 0, \Gamma\ne0$.
Finally,  we demonstrate with case (4) how this decoherent dynamics
can be manipulated with the help of an externally applied
time-dependent control field and set $s\ne0, g\ne 0$ and $\Gamma\ne 0$.

In order to preserve the coherent evolution of the two-level atom and 
to protect it as far as possible from the decoherent influence of the 
environment, we choose the following control scheme: Guided by the 
physics of a rotating wave approximation for the driven system that most 
closely retains the unperturbed dynamics of an initial superposition state 
(\ref{inistate}) we choose the following parameters:  The frequency and the
amplitude of the driving field are
taken to be {\it in resonance} with the level spacing of the two-level system,
i.e.,
$\omega_{\rm L} = \Delta_0$ and  $s=\Delta_0 $ which corresponds to  
a moderately strong driving strength. This choice implies for the ratio of the
corresponding Rabi-frequency and driving strength the value $0.5$. This 
indicates that the rotating wave approximation should be  
used already with caution. 
Note that under the CDT-condition in (\ref{cdtbessel}), the field 
strength would assume an even larger value of $s=2.4048 \Delta_0$.
For the strength of the coupling between the two-level atom and the
cavity mode
we assume $g = 0.05 \Delta_0$. 
This value is consistent  with the range of validity of the
Bloch-Redfield formalism in Born approximation (see above). 
The dissipative system-bath mechanism
is specified as follows: the frequency of the cavity mode  is chosen
to be  in resonance as well, i.e,  $\Omega = \Delta_0$. By doing so, we
 in essence  maximize the
influence of the bath. For the line width of the cavity mode we set 
$\Gamma=0.1 \Delta_0$.
This rather large value mimics (on purpose) an extreme situation because
the line width in most realistic situations is in general much smaller:
Nevertheless, such smaller values would
intensify our appealing  finding of a driving-induced,
enhanced {\it recovery of coherence} even more. Moreover, the temperature
is always set to $T=0$.

Our novel results are depicted in the Fig.\  2 (a)-(c) and the Fig.\ 3 (a)-(c).
Fig.\ \ref{fig2}a depicts the time-evolution of the x-component $\sigma_x(t)$.
The isolated two-level dynamics
 (dashed line) shows coherent oscillations between -1 and 1 at the
frequency of the level spacing $\Delta_0$.
On top of this line one finds (barely visible dotted line)
the results for the driven two-level 
dynamics. This good agreement follows also from the
corresponding rotating wave approximation,
yielding for this preparation just the undriven result. The decoherence
 in presence of a finite
bath coupling ($g= 0.05 \Delta_0,\Gamma=0.1 \Delta_0$), 
see the dashed-dotted line, yields
 an oscillatory decay towards equilibrium $\sigma_x(t\rightarrow \infty) = 0$, 
 whose envelope is made visible by the connecting solid line.
 Next we switch on the cw-laser control
field. As a main result we find that the decoherence becomes considerably
slowed
down -- following closely the isolated driven dynamics. This enhanced
recovery
of coherence for the dissipative {\it driven} dynamics is made visible
to the
eye by the connecting weakly decaying and oscillating envelope. This surprising
result is rooted in the following facts:  
The dissipative, non-driven dynamics experiences
a most  effective dissipation. This is due to the resonant coupling 
at $\Omega=\Delta_0$ of the two-level atom to bath with 
the effective spectral density in (\ref{jeff}) which peaks at $\omega=\Omega$.  
 In contrast, the strong
driving now dresses this level spacing, and moves it out of resonance with 
 the lossy cavity mode. This results in a considerable slow down of driven 
decoherence for $\sigma_x(t)$.

The decoherent dynamics of the $y$-component $\sigma_y(t)$
is qualitatively similar to
$\sigma_x(t)$. It is depicted in Fig.\ \ref{fig2}b for the same choice of
 parameters.

The population difference $\sigma_z(t)$ is shown in Fig.\ \ref{fig2}c.
For the isolated two-level dynamics, $\sigma_z(t)$ remains constant at zero 
(dashed line) since
the system is in an equally weighted superposition
of two eigenstates, yielding an obvious  zero population difference.
In presence of the cw-laser control field, the driven dynamics (dotted line)
yields a finite oscillation of population difference. This deviation from zero
also reflects the deviation from the corresponding rotating wave solution
(being identically zero for this preparation). Nevertheless, this driven
dynamics still exhibits an approximate periodicity
 that closely coincides with the Rabi-value $\Omega_R = \Delta_0/2$.

The undriven, dissipative
relaxation to equilibrium (dashed-dotted line) proceeds with temperature $T=0$
 almost completely
towards the ground state with corresponding
maximal population difference  $\sigma_z(t\rightarrow \infty) \approx 1$.
  Due to the coupling to the cavity mode performing
zero point oscillations, the value of 1 is not fully reached.
The driven, dissipative relaxation (solid line) to the time periodic
asymptotic state  exhibits  oscillations around zero --
following initially (up to $\Delta_0 t \approx 50$)
  closely  the driven coherent dynamics. 
In virtue of Floquet theory for the long-time limit of the time-periodic
generalized Bloch-Redfield equations in (11), this asymptotic periodicity 
matches  in the lon-time limit the
 frequency of driving, i.e.,  $\omega_{L} =\Delta_0$ (not depicted). 
%
%

\subsection{Controlling the decoherence from the atom ground state}
\label{results2}
To answer the question whether the proposed control scheme works as well
in the opposite limit of 
an initial state which is an eigenstate  we next choose  the ground state
 as the initial preparation, i.e., we use
$\sigma_x(t=0)= \sigma_y(t=0)=0, \sigma_z(t=0)=1$. The remaining parameters are
taken to be the same as in the previous subsection \ref{results}.

Fig.\ \ref{fig3}a shows the decoherent dynamics for $\sigma_x(t)$. Since the
chosen initial state is an eigenstate of the isolated two-level system
no dynamics is exhibited (note the filled squares on the line at zero
 in the figure). This situation remains unaltered in presence of a dissipative
coupling of the quantum system, as indicated by the asterisks 
on the line at zero. At zero temperature the system at weak dissipation
 remains essentially in its ground state.
 Upon switching on the driving with no coupling to the lossy cavity present,
 the driven
two-level dynamics  exhibits a Rabi-like quasiperiodic,
 oscillatory behaviour (dotted line). This nonperiodic behavior is rooted
in the
deviation of the full Floquet dynamics from a rotating wave prediction. With
our strong driving strength we {\it a priori} cannot expect good agreement
with
the corresponding rotating wave approximation.
 The coupling to the lossy cavity mode  damps this quasiperiodic  behaviour, 
 following for short times the driven isolated dynamics (see solid line), 
 before
settling down to asymptotic, long-time oscillations at the frequency
 of driving $\omega_{\rm L}=\Delta_0$  
 with a finite, but strongly reduced amplitude (not depicted).

The decoherent dynamics of the $y$-component $\sigma_y(t)$ is again
qualitatively similar to that of 
$\sigma_x(t)$. It is presented in Fig.\ \ref{fig3}b for the same set of
coupling and driving  parameters.

Finally, the time evolution of the population difference $\sigma_z(t)$ is depicted
with Fig.\ \ref{fig3}c.
Clearly, the isolated dynamics from a prepared initial ground
state remains constant at
 $\sigma_z(t) = 1$
(filled squares). The driven dynamics of the two-level system
 exhibits
 strong non-detuned Rabi oscillations  at frequency $ \Omega_R = \Delta_0 / 2$
 between -1 and 1 (dotted line). In this case the rotatating wave prediction
(not depicted)
actually yields  surprisingly good qualitative 
agreement with the exact dynamics.

The case of no driving ($s=0$) but with a coupling to the bath 
($g=0.05 \Delta_0, \Gamma=0.1\Delta_0$) shows again
a trivial dynamics. It relaxes in this case of weak dissipation
with a small relaxation rate towards a slightly reduced constant value
close to 1 (indicated by the asterisks).

The case with resonant driving ($s=\Delta_0, \omega_{\rm L} =\Delta_0$) 
switched on and simultaneous coupling to the lossy
 cavity mode (with $g=0.05 \Delta_0, \Gamma=0.1\Delta_0$)  
 exhibits damped Rabi-oscillations (solid line); it
 eventually settles down in the asymptotic long-time limit
to periodic asymptotic oscillations at twice the
Rabi frequency  and amplitude smaller than 1 (not depicted).

\section{Conclusions} \label{conclusio}
In this work  we have investigated the possibility to control the
time-evolution
of a two-level atom by time-dependent external, periodic control forces. We
have  demonstrated that the
coherent dynamics of the system can be brought to an almost perfect
standstill by choosing
the ratio  of driving amplitude $s$ and driving frequency $\omega_{\rm L}$
at a zero of the Bessel function $J_0(s/\omega_{\rm L})$
({\em coherent destruction of tunneling}).
For an initially prepared
quantum superposition of states all three components $\sigma_i,i=x,y,z$ 
and therefore the entire density matrix $\hat{\rho}$ can
 be locked simultaneously. For the initially prepared ground-state, the
x-component
$\sigma_x$ can be stabilized; the other two components
$\sigma_y$ and $\sigma_z$, however, depict strong (non-Rabi) oscillations.

In presence
of decoherence in a lossy cavity we illustrate that the atomic states can
be dressed by a time-dependent force which moves the atom and the cavity mode
out of resonance. As a consequence, decoherence becomes strongly suppressed. We
have illustrated this
effect for two different initial preparations of the atom: (i) for
 a quantum superposition of states we show that the decoherence
can be suppressed efficiently.  (ii) The second preparation uses the
ground-state wave function of the isolated system.
 In that case the decoherence may also
be slowed down, but the decohering dynamics never approaches again the
initial state.

These findings put the idea across that 
the method can be used to bring back the state of the atom
close to its initial preparation. For the case (i) of a superposition state 
as initial state the decoherent dynamics of the $x$- and $y$-component 
$\sigma_x,\sigma_y$ are similar to the undriven dynamics of the isolated 
two-level system (qubit). Even the $z$-component $\sigma_z$ of the 
driven dissipative dynamics matches 
at distinct instants of time the undriven non-dissipative dynamics. 
For the second case (ii) of the ground state as initial state this idea,  
however, seems to fail for the $z$-component $\sigma_z$.


 To summarize, our proposed  scheme  for controlling the coherent and 
 decoherent  
dynamics of a two-level atom works very well for initially
prepared quantum superpositions of states. This presents good news for the
manipulations of quantum bits (two-level systems) being in
a superposition of states. It is this very feature which makes quantum
computation
interesting and superior to classical computation.
\section*{Acknowledgement}
This work has been supported by the Deutsche Forschungsgemeinschaft
 within the Schwerpunktsprogramm {\it Zeitab\"angige Ph\"anomene und
Methoden in Quantensystemen der Physik und Chemie},
HA1517/14-3 (L.H., I.G., P.H.),
 within the Schwerpunktsprogramm {\it Quanten-Informationsverarbeitung}
 HA1517/19-1, (M.T., P.H.) and in part by the Sonderforschungsbereich
 486 of the Deutsche Forschungsgemeinschaft (I.G., P.H.).
%


\newpage 


\setcounter{figure}{0}

\begin{center}
{\large FIGURES}
\end{center}

\begin{figure}[t]
\centerline{\hbox{
\epsfig{figure=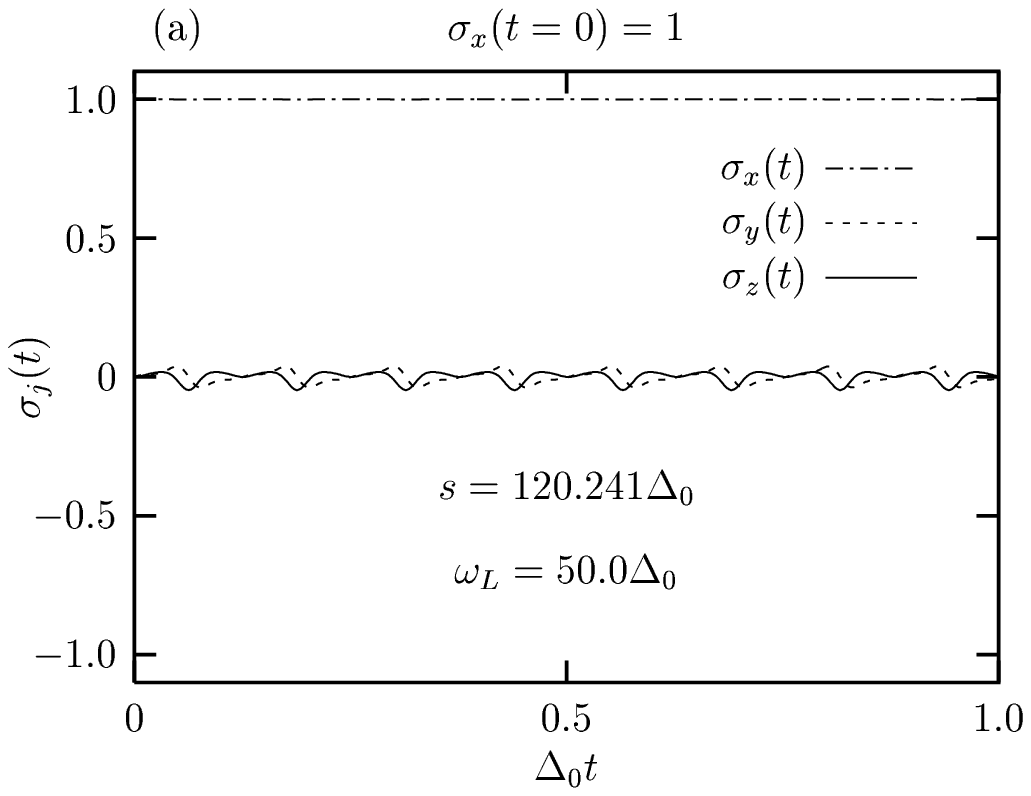,width=95mm,height=95mm,angle=0}
\hfill\epsfig{figure=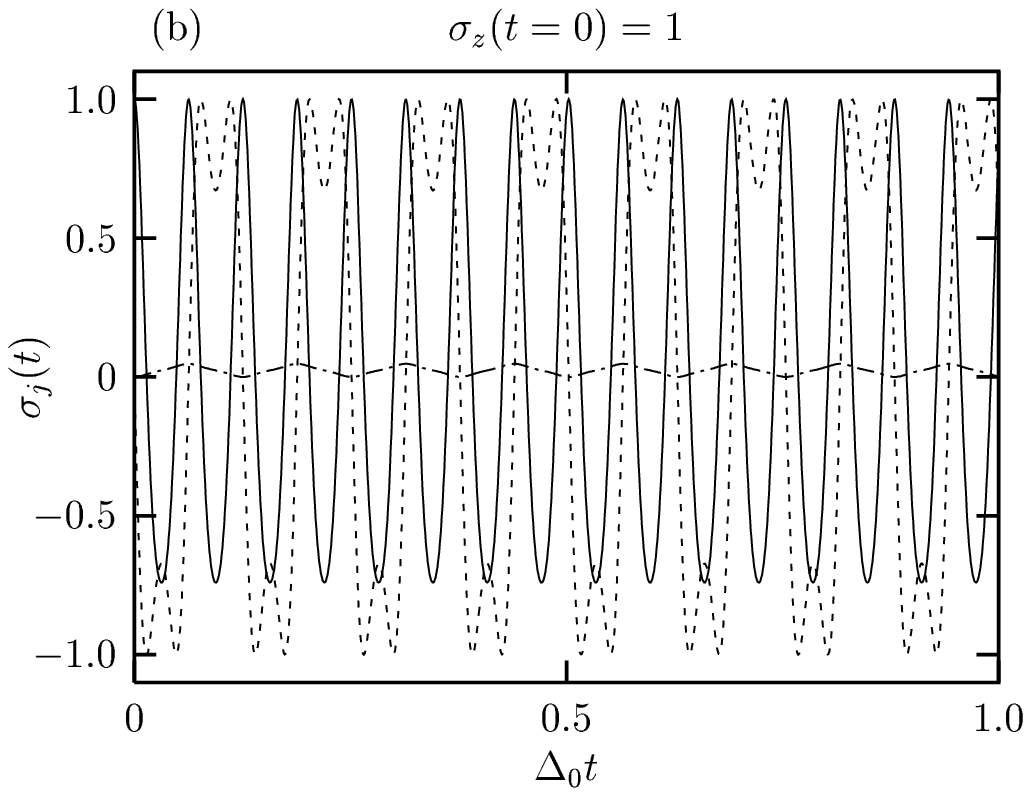,width=95mm,height=95mm,angle=0}
}
}
\caption{Fig.\ \ref{fig1}a: Time-dependence of the expectation values
$\sigma_{j}(t), j=x,y,z$,
determining the dynamics of the 
full density matrix of the two-level atom according to
(\ref{twobytwo},\ref{heisi}). The driving parameters are chosen such that the
condition (\ref{cdtbessel}) for coherent destruction of tunneling is
fulfilled, i.e., the driving amplitude is set to $s=120.241 \Delta_0$ and
the driving frequency $\omega_{\rm L}=50 \Delta_0$.
The two-level system is prepared in an equally weighted
 superposition of the two energy eigenstate, i.e., 
  $\sigma_x(t=0)=1, \sigma_y(t=0)=\sigma_z(t=0)=0$. Fig.\ \ref{fig1}b:
The same as in Fig.\ \ref{fig1}a but for an
initial state preparation 
being the ground state of the two-level system, i.e.,
  $\sigma_x(t=0)= \sigma_y(t=0)=0, \sigma_z(t=0)=1$.
  \label{fig1}}
\end{figure}

\begin{figure}[t]

\newpage

\centerline{\hbox{
\epsfig{figure=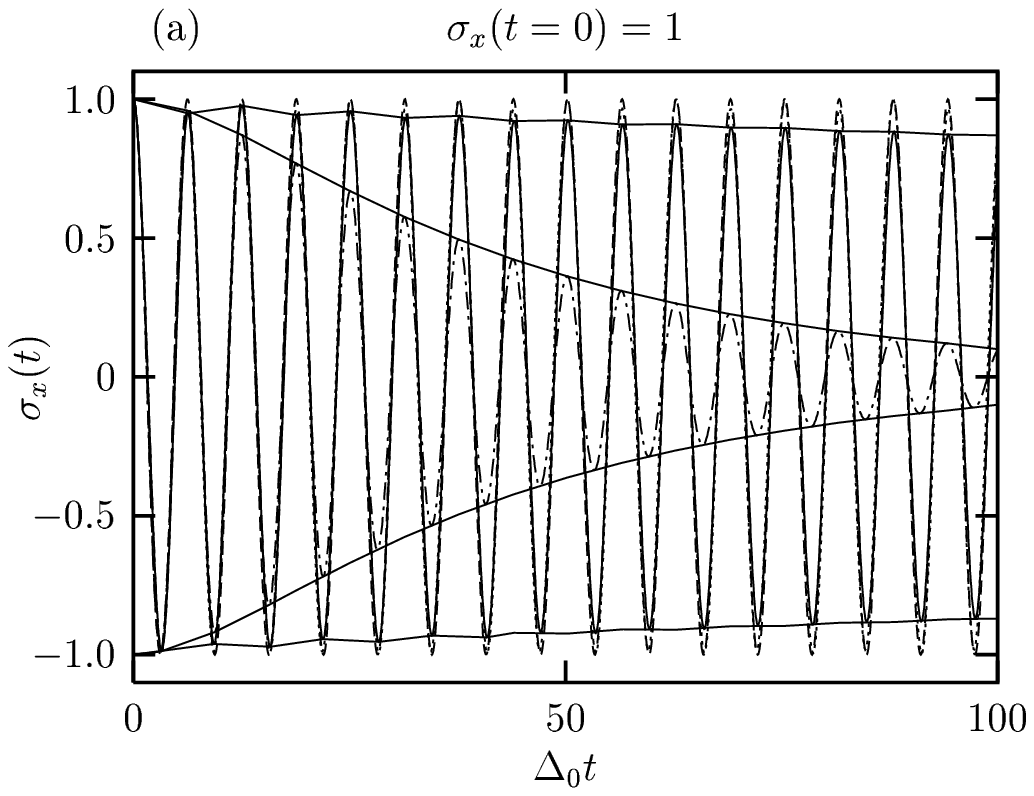,width=95mm,height=95mm,angle=0}
\hfill
\epsfig{figure=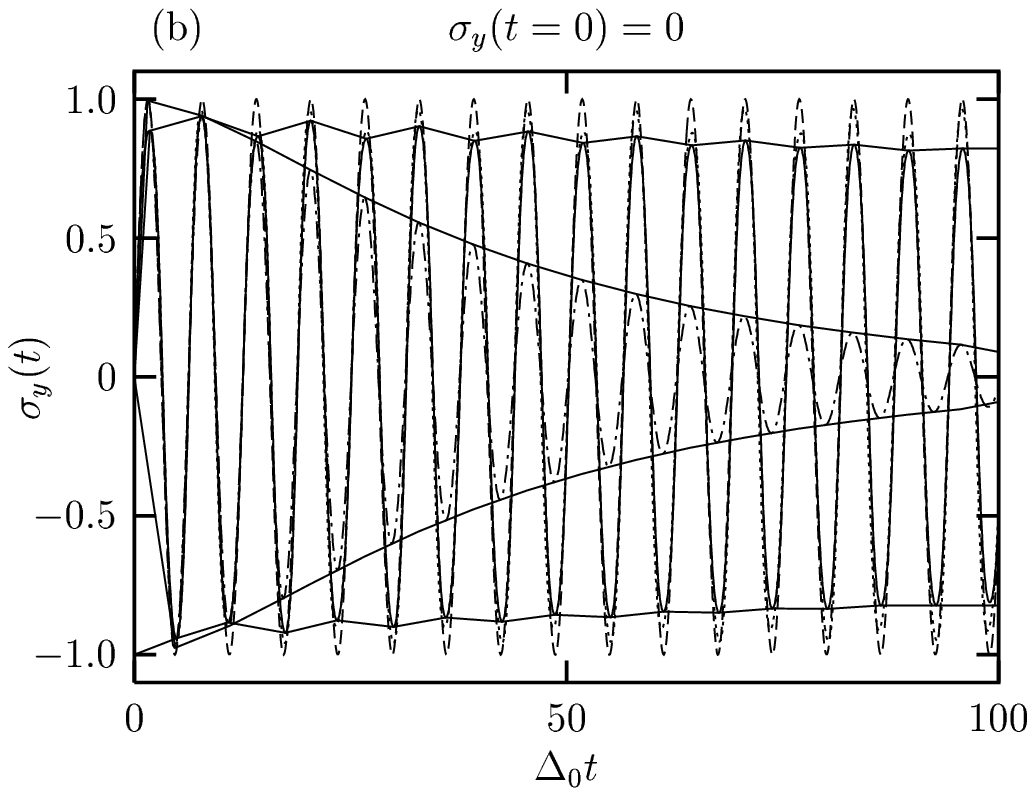,width=95mm,height=95mm,angle=0}
}
}
\centerline{
\epsfig{figure=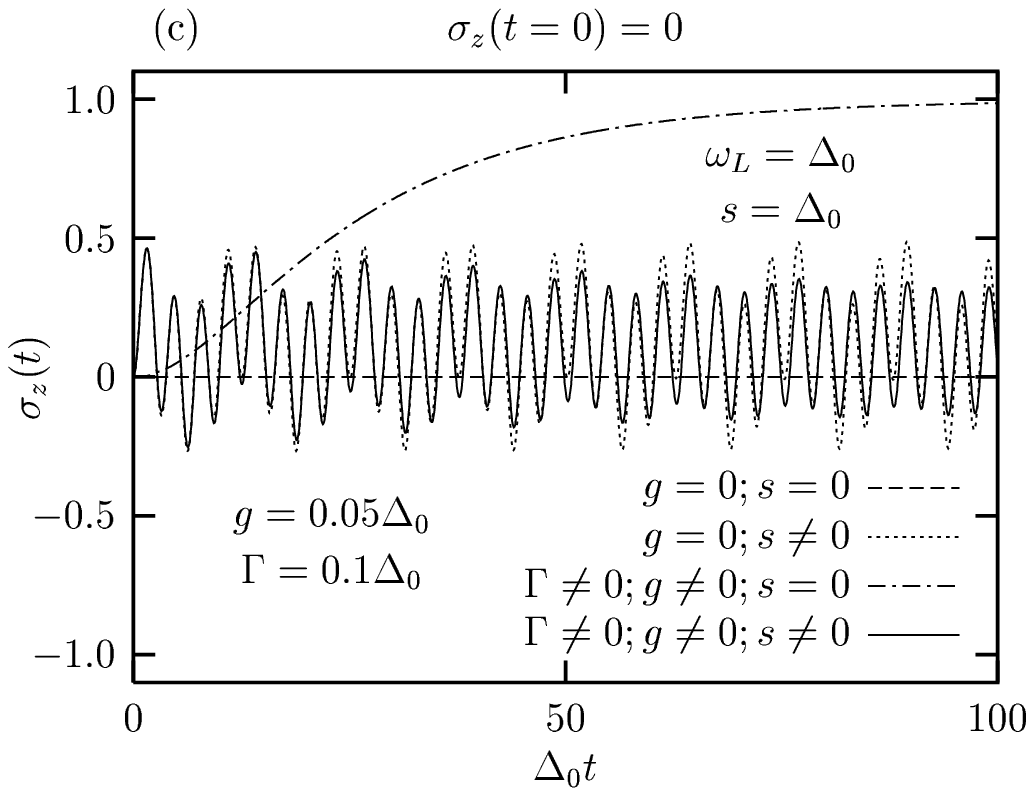,width=95mm,height=95mm,angle=0}
}
\caption{Time-dependence of the expectation values
$\sigma_{x}(t)$ (Fig.\ \ref{fig2}a), $\sigma_{y}(t)$ (Fig.\ \ref{fig2}b) and
$\sigma_{z}(t)$ (Fig.\ \ref{fig2}c)
for the two-level atom initially prepared 
in an equally weighted superposition of the energy eigenstates, i.e.\
$\sigma_x(t=0)=1, \sigma_y(t=0)=\sigma_z(t=0)=0$. Shown are four
cases: (1) no driving ($s=0$), zero system-cavity mode coupling
($g=0$)
(dashed line), (2) with  resonant  driving
($s=\Delta_0, \omega_{\rm L}=\Delta_0$), but zero
coupling  ($g=0$)(dotted line), (3) zero driving ($s=0$), but with finite
coupling
($g=0.05 \Delta_0, \Gamma=0.1 \Delta_0$) 
(dashed-dotted line) and (4) with resonant driving 
($s=\Delta_0, \omega_{\rm L}=\Delta_0$) and with coupling
($g=0.05\Delta_0, \Gamma=0.1 \Delta_0$) (full line). 
The temperature is chosen to be $T=0$ and the cavity-mode frequency is 
set to $\Omega = \Delta_0$.  As a guide for the eye, 
we mark the envelope of the decaying oscillations by  solid lines 
in Fig.\ \ref{fig2}a and Fig.\ \ref{fig2}b.
  \label{fig2}}

\end{figure}

\newpage

\begin{figure}[t]
\centerline{\hbox{
\epsfig{figure=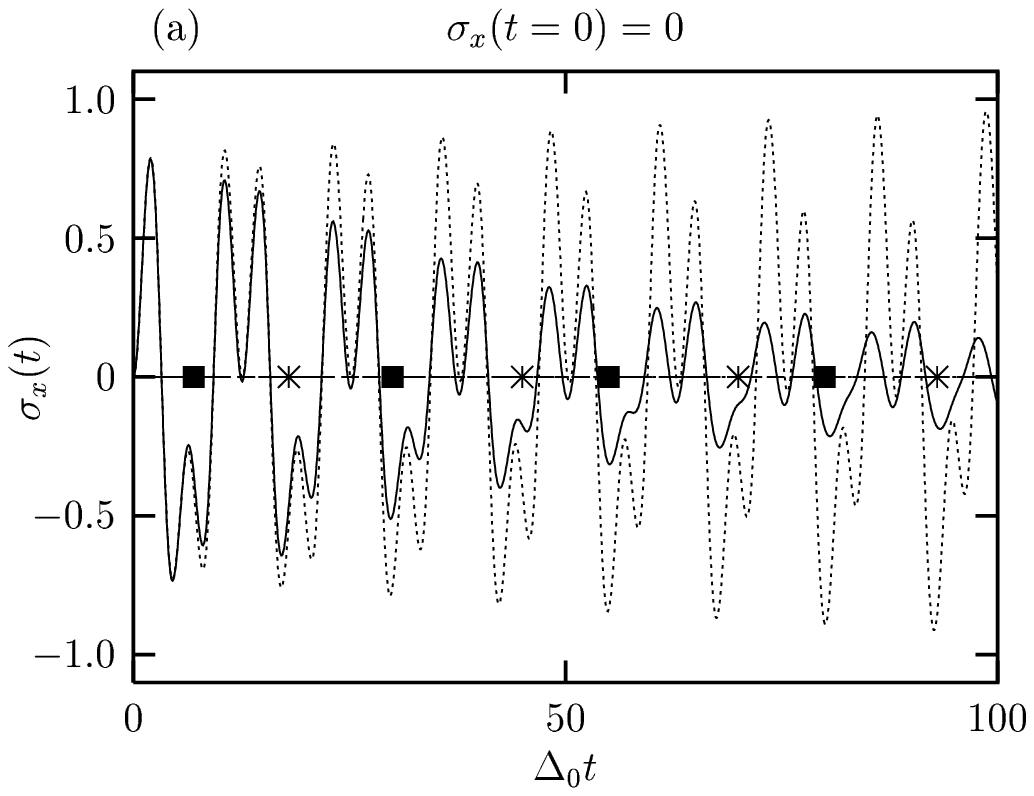,width=95mm,height=95mm,angle=0}
\hfill
\epsfig{figure=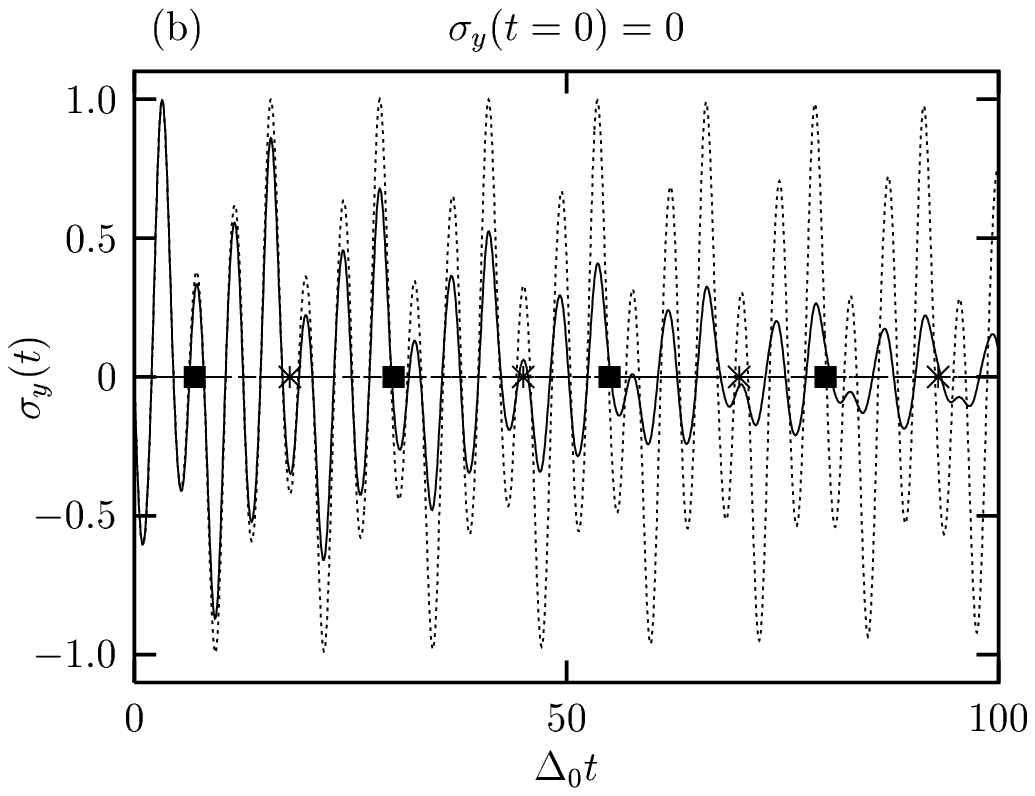,width=95mm,height=95mm,angle=0}
}
}
\centerline{
\epsfig{figure=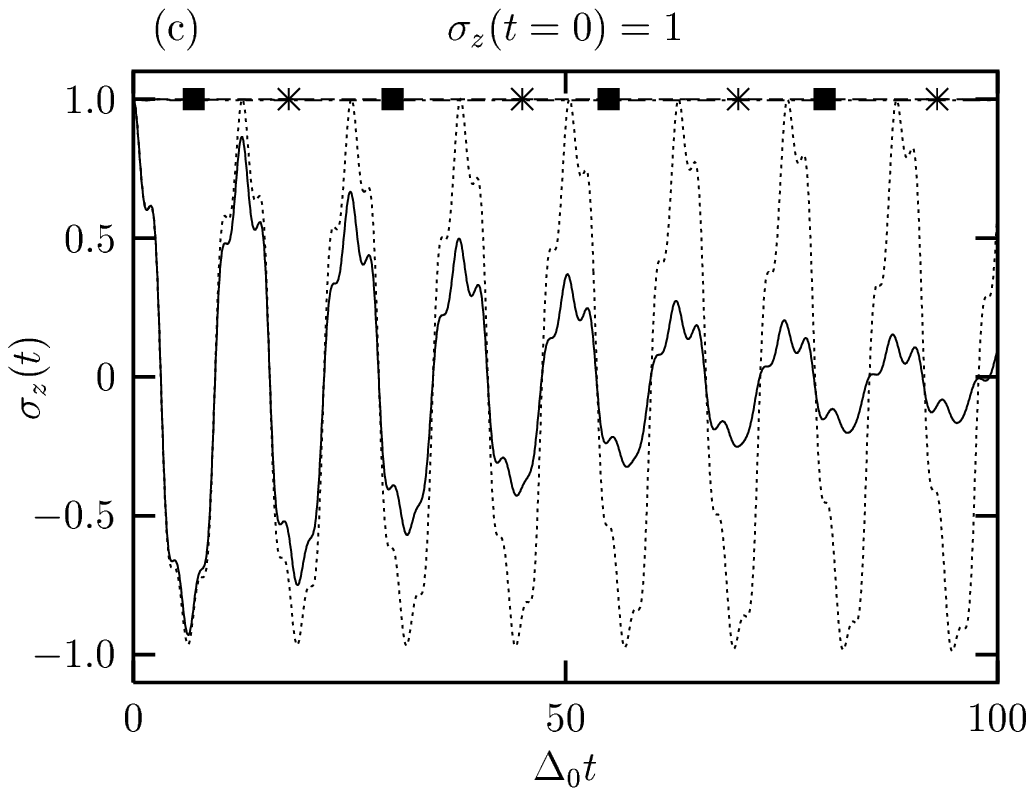,width=95mm,height=95mm,angle=0}
}
\caption{Time-dependence of the expectation values
$\sigma_{x}(t)$ (Fig.\ \ref{fig3}a), $\sigma_{y}(t)$ (Fig.\ \ref{fig3}b) and
$\sigma_{z}(t)$ (Fig.\ \ref{fig3}c) for the two-level atom initially prepared 
in the ground-state, i.e.,
$\sigma_x(t=0)= \sigma_y(t=0)=0, \sigma_z(t=0)=1$. Similar to Fig.\
\ref{fig2}
four cases are depicted: (1) no driving ($s=0$), isolated 
two-level system ($g=0$) (dashed line with filled squares $\blacksquare$), 
(2) with  resonant driving 
($s=\Delta_0, \omega_{\rm L}=\Delta_0$), no dissipation
($g=0$)(dotted line), (3) no driving ($s=0$), with dissipation
($g=0.05\Delta_0, \Gamma=0.1 \Delta_0$) 
(dashed-dotted line with asterisks $\ast$) 
and (4) with resonant driving 
($s=\Delta_0, \omega_{\rm L}=\Delta_0$) and
with dissipation ($g=0.05\Delta_0, \Gamma=0.1 \Delta_0$) 
(full line). The temperature is chosen to be $T=0$ and 
and the cavity-mode frequency is set to $\Omega = \Delta_0$.
  \label{fig3}}

\end{figure}

\end{document}